\pdfoutput=1
\documentclass[%
 reprint,
  amsmath,amssymb,
   aps,
   prd,
   ]{revtex4-1}
\usepackage[english]{babel}
\usepackage{epsfig,array}
\usepackage{multirow,xcolor}
\usepackage{hyperref}

\usepackage{booktabs}

\newcommand{\be}{\begin{eqnarray}}
\newcommand{\ee}{\end{eqnarray}}

\newcommand{\mpm}{\!\pm\!}

\newcommand{\er}[2]{
  \ensuremath{
  ^{+#2}_{-#1}
  }
}          

\begin{document}

\title{Partial-wave analysis of $J/\psi\to K^+K^-\pi^0$}
\author{ 
\begin{small}
\begin{center}
M.~Ablikim$^{1}$, M.~N.~Achasov$^{10,d}$, P.~Adlarson$^{59}$, S. ~Ahmed$^{15}$, M.~Albrecht$^{4}$, M.~Alekseev$^{58A,58C}$, A.~Amoroso$^{58A,58C}$, F.~F.~An$^{1}$, Q.~An$^{55,43}$, Y.~Bai$^{42}$
% ...
\\
\vspace{0.2cm}
(BESIII Collaboration)\\
\vspace{0.2cm} {\it
$^{1}$ Institute of High Energy Physics, Beijing 100049, People's Republic of China\\
% ...
$^{4}$ Bochum Ruhr-University, D-44780 Bochum, Germany\\
% ...
$^{10}$ G.I. Budker Institute of Nuclear Physics SB RAS (BINP), Novosibirsk 630090, Russia\\
% ...
$^{15}$ Helmholtz Institute Mainz, Johann-Joachim-Becher-Weg 45, D-55099 Mainz, Germany\\
% ...
$^{42}$ Southeast University, Nanjing 211100, People's Republic of China\\
$^{43}$ State Key Laboratory of Particle Detection and Electronics, Beijing 100049, Hefei 230026, People's Republic of China\\
% ...
$^{55}$ University of Science and Technology of China, Hefei 230026, People's Republic of China\\
% ...
$^{58}$ (A)University of Turin, I-10125, Turin, Italy; (B)University of Eastern Piedmont, I-15121, Alessandria, Italy; (C)INFN, I-10125, Turin, Italy\\
$^{59}$ Uppsala University, Box 516, SE-75120 Uppsala, Sweden\\
% ...
$^{d}$ Also at the Novosibirsk State University, Novosibirsk, 630090, Russia\\
% ...
}\end{center}

\vspace{0.4cm}
\end{small}
}

\begin{abstract}
A partial-wave analysis of the decay $J/\psi \to K^+K^-\pi^0$ has been made using
$(223.7\pm1.4)\times 10^{6}$ $J/\psi$ events collected with the BESIII
detector in 2009. The analysis, which is performed within the
isobar-model approach, reveals contributions from $K^*_2(1430)^\pm$, $K^*_2(1980)^\pm$
and $K^*_4(2045)^\pm$ decaying to $K^\pm\pi^0$.
The two latter states are observed in  $J/\psi$ decays for
the first time.
Two resonance signals decaying to $K^+K^-$ are also observed.
These contributions can
not be reliably identified and their possible interpretations
are discussed. The measured branching fraction $B(J/\psi \to K^+K^-\pi^0)$ of 
$(2.88\pm0.01\pm0.12)\times10^{-3}$ is more precise than previous results. Branching fractions for the
reported contributions are presented as well. The results of the partial-wave analysis
differ significantly from those previously obtained by BESII and BABAR.
\end{abstract}

\maketitle

\section{Introduction}

A good knowledge of the spectrum and properties of hadrons is one of
the key issues for understanding the strong interaction at low
and intermediate energies. The conventional quark model implies that
quark-antiquark states are produced as nonets, which consist of
mesons with strange and non-strange quarks. Therefore, an
accurate identification of mesons with one strange quark can
help to establish nonet members in the isoscalar sector, where the
situation is more complicated. This is due to a potential mixing between octet and
singlet states as well as possible mixing with glueball states.

The identification of meson radial excitations  also helps in the 
understanding of quark-antiquark interaction at intermediate energies.
Quark potential models~\cite{Godfrey:1985xj} predict that the squared masses of
radial excitations depend on the excitation number quadratically.
However, in the analysis of proton-antiproton annihilation in flight,
it was found that this dependence is close to the linear one similar
to the Regge trajectories~\cite{Anisovich:2000kxa}.
If correct, this behavior has the potential to reveal a new symmetry of
the quark-antiquark interaction
\cite{Afonin:2013npa,Afonin:2014nya}. Therefore, the experimental confirmation
(or disproof) of this behavior is an important task
in experimental hadron physics.

$J/\psi$ decays are ideal for the study of meson spectra and
the determination of meson properties. They can
provide important information about meson states with masses up to
3~GeV/$c^2$ and partial-wave analysis  is facilitated due to
the well-known
quantum numbers of the initial state. Moreover, the $J/\psi$
radiative decay is favored for the production of glueball states
which makes it a perfect tool to search for and study such exotics \cite{Klempt:2007cp}.

In this paper we report the results of a partial-wave analysis
(PWA) of the decay $J/\psi \to K^+K^-\pi^0$.
This decay channel has been previously studied by the
MARK \cite{Vannucci:1976qg}, MARK-II \cite{Franklin:1983ve},
MARK-III \cite{Coffman:1988ve}, 
DM2 \cite{Jousset:1988ni}, BESII \cite{Ablikim:2006hp}, and BABAR \cite{Aubert:2007ym,Lees:2017ndy} Collaborations, but only two recent
publications report PWA results.
In the first of these~\cite{Ablikim:2006hp}, BESII analyzes
58~million $J/\psi$ decays and observes
a very broad exotic resonance $X(1575)$
with pole position $\left[ ( 1576^{+49}_{-55}\,^{+98}_{-91}) -
i(409^{+11}_{-12}\,^{+32}_{-67}) \right] \,{\rm MeV}/c^2$
and branching fraction $B(J/\psi \to X(1575)\pi^0 \to K^+K^-\pi^0)=
\left(8.5\pm0.6_{-3.6}^{+2.7}\right)\times10^{-4}$. In the second
analysis~\cite{Lees:2017ndy}, BABAR reports a PWA solution
based on a smaller data set of 2102 events, which
consists of $K^*(892)^\pm$, $K^*(1410)^\pm$ and $K_2^*(1430)^\pm$ states in the
$K^\pm\pi^0$ channels, while the enhancement at low $K^+K^-$ invariant masses is attributed to the $\rho(1450)$.
The analysis presented in this paper is based on a data set of 182,972 event candidates selected from $(223.7\pm1.4)\times10^6$  $J/\psi$
decays \cite{Ablikim:2016fal} collected by
the BESIII experiment in 2009. 
The high statistics and  good data quality allow us to reveal
signals from states that have not been observed before and
precisely determine properties of intermediate states.
Moreover, the obtained PWA
solution can be used for the simulation of the
irreducible background from this channel to the $J/\psi \to \gamma K^+K^-$
decay, which is one of the key channels to be studied in the  search for a
low-mass glueball.

\section{BESIII experimental facility}
The BESIII detector is a magnetic
spectrometer~\cite{Ablikim:2009aa} located at the Beijing Electron
Positron Collider (BEPCII)~\cite{Yu:IPAC2016-TUYA01}. The
cylindrical core of the BESIII detector consists of a helium-based
 multilayer drift chamber (MDC), a plastic scintillator time-of-flight
system (TOF), and a CsI(Tl) electromagnetic calorimeter (EMC),
which are all enclosed in a superconducting solenoidal magnet
providing a 1.0~T magnetic field. The solenoid is supported by an
octagonal flux-return yoke with resistive plate counter muon
identifier modules interleaved with steel. The geometrical acceptance of
charged particles and photons is 93\% over the $4\pi$ solid angle. The
charged-particle momentum resolution at $1~{\rm GeV}/c$ is
$0.5\%$, and the $dE/dx$ resolution is $6\%$ for electrons
from Bhabha scattering. The EMC measures photon energies with a
resolution of $2.5\%$ ($5\%$) at $1$~GeV in the barrel (end cap)
region. The time resolution of the TOF barrel part is 68~ps, while
that of the end-cap part is 110~ps.

The {\sc geant4}-based simulation software BOOST \cite{BOOST}
is used to simulate the detector response.
An inclusive $J/\psi$ Monte Carlo (MC) sample is used to estimate
the background. In this sample the production of the $J/\psi$ resonance
is simulated by the MC event generator KKMC \cite{Jadach:1999sf, Jadach:2000ir}
and decays are generated by {\sc evtgen} \cite{Lange:2001uf, Ping:2008zz}.
The branching fractions of  known decay modes  are set to the
Particle Data Group (PDG) \cite{Amsler:2008zzb} world-average values and
the remaining unknown decays are generated according to the Lund-Charm
model \cite{Chen:2000tv}.

\section{Event selection}
The $K^+K^-\pi^0$ candidate events are required to have two charged tracks
with zero net charge and at least two good photons.

Charged tracks must be reconstructed within the geometrical acceptance
of the detector ($|\cos\theta|<0.93$, where $\theta$ is the
angle with respect to the beam axis)
 and originate from the interaction point
($|z|<10$~cm and $R<1$~cm, where $z$ and $R$ are minimal distances from
a track to the run-averaged interaction point along the beam direction and
in the transverse plane, respectively).
An event is rejected if the transverse momentum of at least one charged track
is too low ($p_T<120$ MeV/$c$).
Particle identification (PID) is performed using
TOF and MDC $dE/dx$ information. Their measurements are combined
to form particle identification confidence levels (C.L.) for  $\pi$, $K$,
and $p$ hypotheses, and the particle type with the highest C.L. is assigned
to the track. Both tracks are required  to be identified as kaons.

Signal clusters  in the EMC within the acceptance region, which are not associated with charged tracks and 
possess energy $E>25$~MeV in the barrel part of the detector
and $E>50$~MeV in the end caps,
are treated as photon candidates.
To exclude showers from association with charged particles,
the angle between the shower direction and the
charged tracks extrapolated to the EMC must be greater
than 10 degrees. 
The requirement on the EMC cluster time with respect to the start of 
the event ($0\text{ ns} \le t \le 700$~ns)
is used to reject
electronic noise and energy deposits not related to the analyzed event.

Consistency between the detector response and a final state hypothesis
(for the signal and specific background decays)
is evaluated by a four-momentum constrained (4C) kinematic fit.
Firstly, the accepted pair of charged tracks and each pair of
the selected photon candidates with invariant mass
$M_{\gamma\gamma} < 300$ MeV/$c^2$ are fitted under
the $\gamma\gamma K^+K^-$ hypothesis.
A combination with the lowest value of $\chi^2_{(4C)\gamma\gamma K^+K^-}$
is selected and an event is retained if $\chi^2_{(4C)\gamma\gamma K^+K^-} < 60$.
Secondly, the $\chi^2_{(4C)\gamma\gamma K^+K^-}$ is compared to 
the corresponding value obtained in the best fits under the main
background hypotheses:
$\gamma\gamma\pi^+\pi^-$, $\gamma K^+K^-$, and, in the cases more than two good photon candidates are selected, $\gamma\gamma\gamma K^+K^-$. If
any of the background hypotheses results in a lower $\chi^2$ value,
the event is rejected.
Finally, the $\pi^0$ candidates are reconstructed requiring the two-photon mass
of the selected pair to be within a
$110\text{ MeV/}c^2 < M_{\gamma\gamma} < 150$~MeV/$c^2$
interval.
For the partial-wave analysis, we use particle momenta after the
five-constrained (5C) kinematic fit, which also constrains
the invariant mass of the selected photon pair to the nominal
$\pi^0$ mass.

A total of 182,972 candidates satisfy the selection criteria.
The corresponding number of background events is estimated from
the inclusive MC: $N_{bg}=565\mpm24$ (or 0.3\%).
The largest background contributions come from the decay channels
$J/\psi \to \gamma \eta_c, \, \eta_c \to
K^+K^-\pi^0$ and $J/\psi \to \gamma K^+K^-$.
The continuum background, i.e. that due to the $e^+e^-\to \gamma^* \to K^+K^-\pi^0$
process, is estimated from the analysis of a data sample
of approximately 280~nb$^{-1}$ collected from $e^+e^-$
collisions at $3.08$~GeV.
It gives $N_{continuum}=855\pm499$,
where the uncertainty is statistical.
The background treatment in the PWA will be described in
the next section.

\begin{figure}[pt]
\begin{center}
    \includegraphics[width=0.45\textwidth]{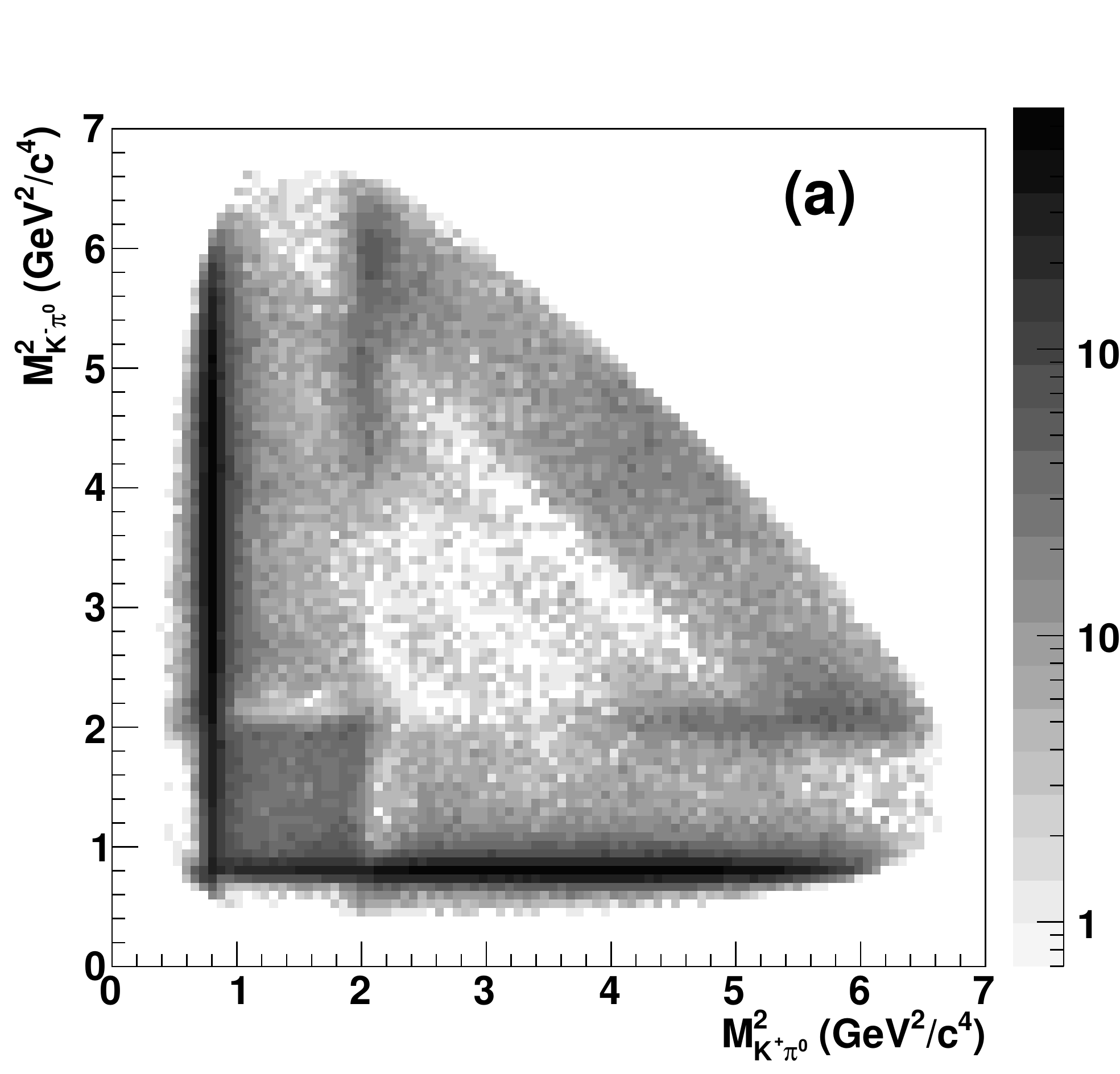}
    \includegraphics[width=0.45\textwidth]{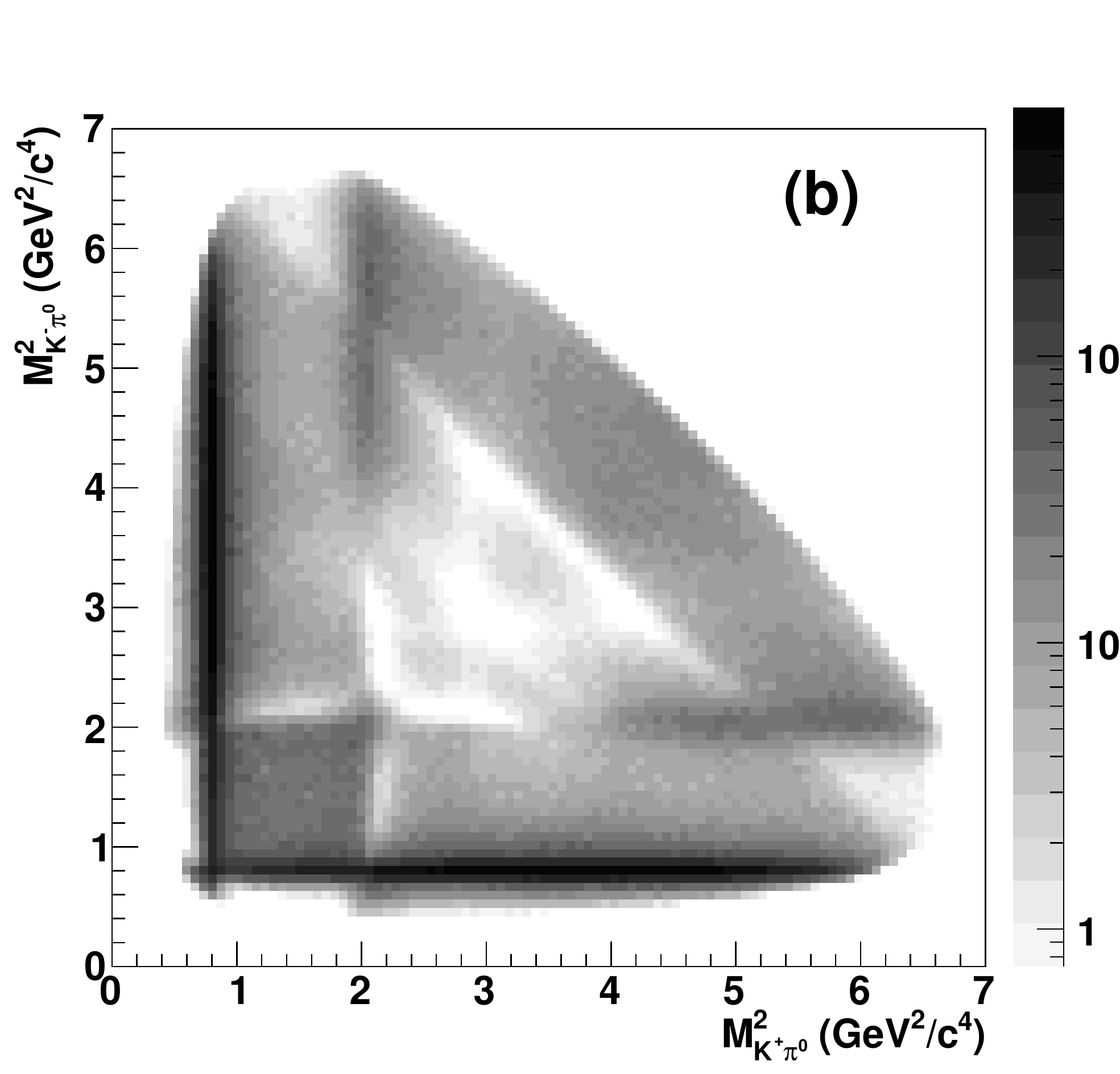}
    \includegraphics[width=0.45\textwidth]{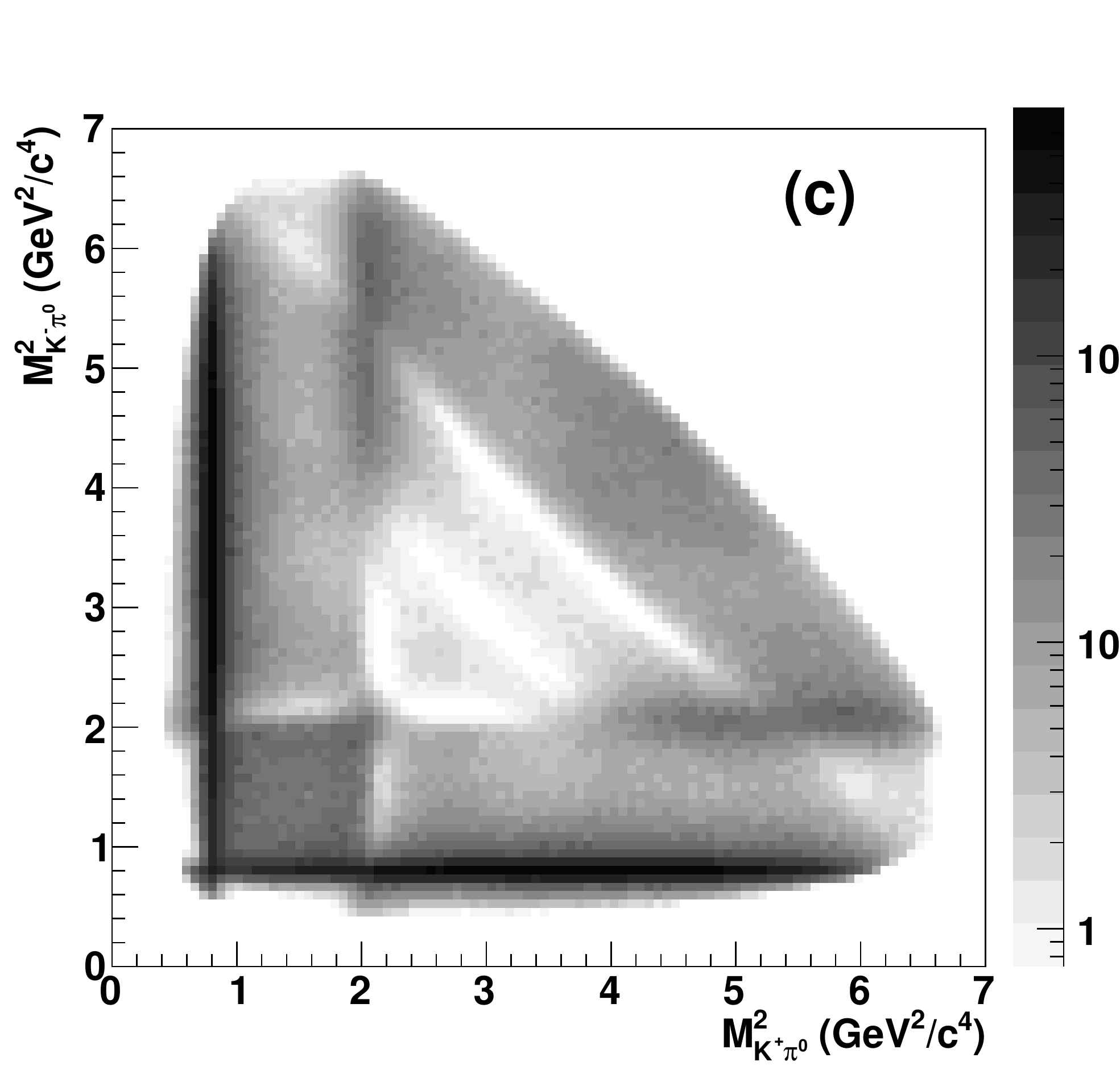}
    \caption{Dalitz plots for the selected data (a), the PWA solution~I (b) and
    the PWA solution~II (c).}
    \label{fig:dalitz}
\end{center}
\end{figure}

The Dalitz plot for the selected data is shown in Fig.~\ref{fig:dalitz}(a).
Its most striking feature is a clear $K^*(892)^\pm$ signal.
In the internal region of the plot a clear signal from $K_2^*(1430)^\pm$
is seen as well as structures at $M^2(K^\pm\pi^0) \approx$ 4~GeV$^2$/$c^4$. These structures are likely to be the result of positive interference of resonances in
the $K^\pm\pi^0$ channels. In the $K^+K^-$ channel there are indications
for a resonance signal at 1.6~--~1.7~GeV/$c^2$ and a signal at
higher masses. 

\section{Partial-wave analysis}
We use the isobar model to describe the $J/\psi$ decay into  $K^+K^-\pi^0$. The amplitude is
parameterized as a sum of sequential quasi two-body decay processes in this approach.
The subprocess described by intermediate state production and the subsequent 
decay to a specific
pair of the final state mesons is referred to as
the decay kinematic channel.
The angular-dependent parts of the partial-wave
amplitudes are calculated in the framework of the covariant tensor
approach as described in detail in Ref.~\cite{Zou:2002ar}.
Note that in our case the conservation of $P$- and $C$-parities
restricts the number of allowed partial waves for 
production and decay of any resonance to one.
To account for the finite size of a hadron
each decay vertex also includes Blatt-Weisskopf form factors,
which depend on the Blatt-Wesskopf radius $r$.
The Breit-Wigner term
for the resonance $a$ in the kinematic channel $m$
(labeled by the number of the spectator particle)
is
\be
\nonumber
A^{BW}_{m,a}\!=\!\!
\frac{1}{M_a^2\!-\!s_m\!-\!iM_a\Gamma(s_m, J_a)}.
\nonumber
\ee
Here $M_a$, $J_a$ and $s_m$ are the resonance mass, spin and
the invariant mass squared of its daughter particles, respectively.
The width of the $K^*(892)^\pm$ state is defined by its decay
to $K\pi$ and is parameterized as:
\be
\nonumber
\begin{split}
	\Gamma(s_m, J_a)=\frac{\rho_J(s_m)}{\rho_J(M_a^2)}\Gamma_a, \\
	\rho_J(s_m)=\frac{2q}{\sqrt{s_m}}\frac{q^{2J}}{F^2(q^2,r,J)}.
	\label{amp}
\end{split}
\ee
Here, $\Gamma_a$ is the resonance width, $q$ is the
relative momentum of the daughter particles
calculated in the resonance rest frame and $F(q^2,r,J)$
is the above-mentioned Blatt-Weisskopf form factor.
The same parameterization is used for the width of
the $K_2^*(1430)^\pm$ resonance, whose decay branching fraction
to $K\pi$ is about 0.5.
For other states we use a constant
width $\Gamma(s_m, J_a) = \Gamma_a$ due to the absence of
reliable information about their branching fractions.

The masses, widths, decay radii (for the $J/\psi$, $K^*(892)^\pm$
and $K_2^*(1430)^\pm$) of resonances as well as
the product of their production and decay couplings
(complex numbers in general case) are initially free parameters
of our fit. We find fit results weakly sensitive to the $J/\psi$
decay radius. Hence, we set this parameter to be $0.7$~fm,
as is obtained in Ref.~\cite{Bugg:1995jq}.

The analysis is performed within the framework of the
event-by-event maximum likelihood method, which allows us to take
into account all correlations in the multidimensional phase space.
The negative log-likelihood function NLL is expressed as 
\be
\text{NLL}=-\sum\limits_i \ln \frac{\omega_i\epsilon_i}{\int \epsilon \omega d\Phi} =
-\sum\limits_i \ln \frac{\omega_i}{\int \epsilon \omega d\Phi}+const
\nonumber
\ee
and is minimized. Here index $i$ runs over the selected data events,
$\omega_i$ is the decay-amplitude squared, summed over
transverse $J/\psi$ polarizations and evaluated from the four-momenta
of final particles in the event $i$.
The detector and event selection efficiency for the measured
four-momenta is denoted by $\epsilon_i$, 
the denominator is a normalization integral over the phase space ($\Phi$),
and the $const$ term is independent of the fit parameters.
The normalization integral is calculated
using phase-space distributed MC events that
pass the detector simulation and the event reconstruction.
To take the background into account we estimate its
contribution to the NLL function and subtract it.
This is done by the evaluation of the NLL function over properly
normalized data samples that have a kinematic distribution
similar to that of the background. We consider two types of
background channels: those producing a peak
at the $\pi^0$ mass in the two-photon invariant-mass distribution
(``peaking'' background) and those exhibiting a smooth shape
below the peak (``non-peaking'' background).
The former is estimated from
$J/\psi \to \gamma \eta_c$, $\eta_c \to \gamma K^+K^-\pi^0$ events selected
under criteria similar to ones of the main event selection,
and the latter is estimated from the
$\pi^0$ mass cut sideband: $190\text{ MeV/}c^2 <M_{\gamma\gamma} < 230$~MeV/$c^2$.

This approach neglects the detector resolution,
which is a good approximation for all resonances except for
the $K^*(892)^\pm$. The MC simulation shows that estimated
bias to the measured width of $K^*(892)^\pm$ is much larger than the
corresponding systematic uncertainty estimated from other sources.
At the same time, this bias is much smaller than the $K^*(892)^\pm$ width,
which allows us to use the approximation proposed in 
Ref.~\cite{Denisenko:2015yoa} to take into account the detector resolution.
Due to the significant computation time, this method is used
only to correct the final PWA results.

The quality and consistency of the obtained solution is
evaluated by the comparison of the mass and angular
distributions of the experimental data and reconstructed
phase-space generated MC events weighted according to the
PWA solution.

The conservation of $P$- and $C$-parities strongly restricts the 
allowed quantum numbers of intermediate states.
In the $K^\pm\pi^0$ channels only resonances with
quantum numbers $I=1/2$, $J^{P}\!=\!1^-, 2^+,3^-,4^+\ldots$ can be
produced.
The reaction is dominated by $K^*(892)^\pm$
production.   
There are two other established vector states which are in the accessible
mass region: $K^*(1410)$ and $K^*(1680)$ \cite{Tanabashi:2018oca}. In the
$2^+$, $3^-$ and $4^+$ partial waves three states are well established:
$K_2^*(1430)$, $K_3^*(1780)$ and $K_4^*(2045)$. 
Possible contributions must also be considered from two observations reported by the LASS Collaboration: a $2^+$ state at 1980 MeV/$c^2$ \cite{Aston:1986jb}
(also claimed to be seen by SPEC~\cite{Tikhomirov:2003gg})
and a $5^-$ state at 2380 MeV/$c^2$ \cite{Aston:1986rm},
which needs confirmation. As for the $K^+K^-$ channel,
the produced resonances are restricted to quantum
numbers $J^{PC}=J^{--}$, where $J=1,3,5\ldots$.
For the strong decays of the $J/\psi$
isospin and $G$-parity conservation requires
$I^G=1^+$. There are two well known isovector resonances in the
$J^{PC}=1^{--}$ sector: the $\rho(1450)$ and $\rho(1700)$, and a set of
observations that needs confirmation: the $\rho(1570)$, $\rho(1900)$
and $\rho(2150)$ (see Ref.~\cite{Tanabashi:2018oca}). In the isovector
$J^{PC}=3^{--}$ sector one can expect the production of the
well known and relatively narrow $\rho_3(1690)$ state. At higher energies
there have been observations of two $J^{PC}=3^{--}$ states: the $\rho_3(1990)$ and
$\rho_3(2250)$. The first isovector $J^{PC}=5^{--}$ state is expected
to have a mass of around 2350 MeV/$c^2$. Such a resonance is observed
 in the analysis of the
GAMS2 data for the reaction $\pi^-p\to \omega\pi^0n$~\cite{Alde:1994jm}
and in the  analyses of proton-antiproton annihilation
in flight into different meson final states (e.g. see
Ref.~\cite{Anisovich:2002su}).
The decay of the $J/\psi$ through a virtual photon does not forbid but
even favors the production of $I^G=0^-$ resonances.
The $J/\psi \to \phi\pi^0$ decay is strongly suppressed \cite{Ablikim:2015mua},
hence the production of excited $\phi$ mesons is expected
to be negligible assuming the absence of strong mixing
of excited $\phi$ and $\omega$ states.
However, the production of excited $\omega$ resonances is possible. The
isovector and isoscalar states can be distinguished in
a combined analysis of the decay under consideration and the $J/\psi$
decay to $K^\pm K^0\pi^\mp$.

\subsection{Fit to the data}

The masses and widths of all states included in the solution
(with the sole exception of the $\rho(770)$) are
initially free fit parameters.
For the well-established $K\pi$ resonances we use results of the LASS
fits to the elastic $K\pi$ scattering amplitudes \cite{Aston:1987ir} as
reference values. The masses and widths of these states are allowed to vary
within $\pm\sigma$ of the LASS measurements
(here $\sigma$ stands for the LASS uncertainty).
If no NLL minimum is found for the mass or width within this
range or the minimum is unstable (with
respect to variations of the PWA solution used for estimation
of systematic errors), the parameter is set to the central
value of the LASS results.
Motivated by the claim of an observation of the $K_2^*(1980)^\pm$ by LASS \cite{Aston:1986jb}
and by Regge trajectories predicting a state at approximately
1.8~GeV/$c^2$ we introduce a second $J^P=2^+$ contribution
with a mass allowed to vary within the 1.75~GeV/$c^2$~--~2.1~GeV/$c^2$ interval.
Two clear resonance-like $K^+K^-$ signals are found
to significantly contribute to the data description in all
fits. The first contribution has a mass of around 1.65~GeV/$c^2$ and
is likely a manifestation of the $\rho(1700)$ or $\omega(1650)$, or interference between the two. 
Note that the parameters of both these states remain highly uncertain.
For the $\rho(1700)$, the PDG quotes the results with the mass
varying roughly from 1540~MeV/$c^2$ to 1860~MeV/$c^2$, which may
indicate the presence of two states. Quark potential models \cite{Godfrey:1985xj}
suggest two resonances close to this mass range: $1^3D_1$ and $3^3S_1$.
This possibility is implied in the interpretation of the fit results.
The second contribution has a mass of around 2.0~GeV/$c^2$~--~2.1~GeV/$c^2$,
close to the mass of the $\rho(2150)$. No limitations on their parameters
are imposed in the fits. For the $\rho(1450)$ the
mass range from 1.3~GeV/$c^2$ up to 1.5~GeV/$c^2$ is studied,
but no NLL minima are found, and so its mass and width
are fixed to the PDG estimates \cite{Tanabashi:2018oca}.

In the analysis we find that the PWA solution can not be saturated
with well-known states included as Breit-Wigner resonances
and constant contributions in the lowest
partial waves. At the same time, the ``missing part'' of the
PWA solution can not be reliably attributed to a single
resonance and mainly manifests itself as a slow changing
background in the $J^P=3^-$ partial wave of the $K^\pm\pi^0$
pairs at high $K^\pm\pi^0$ masses. Below we provide
two solutions constructed with and without the smooth
contribution in this partial wave to demonstrate that the 
conclusions of this analysis are not strongly affected by
assumptions on the ``missing part'' of the PWA solution.

\subsection{Solution I}
\begin{figure}[t]
\centerline{\includegraphics[width=0.48\textwidth]{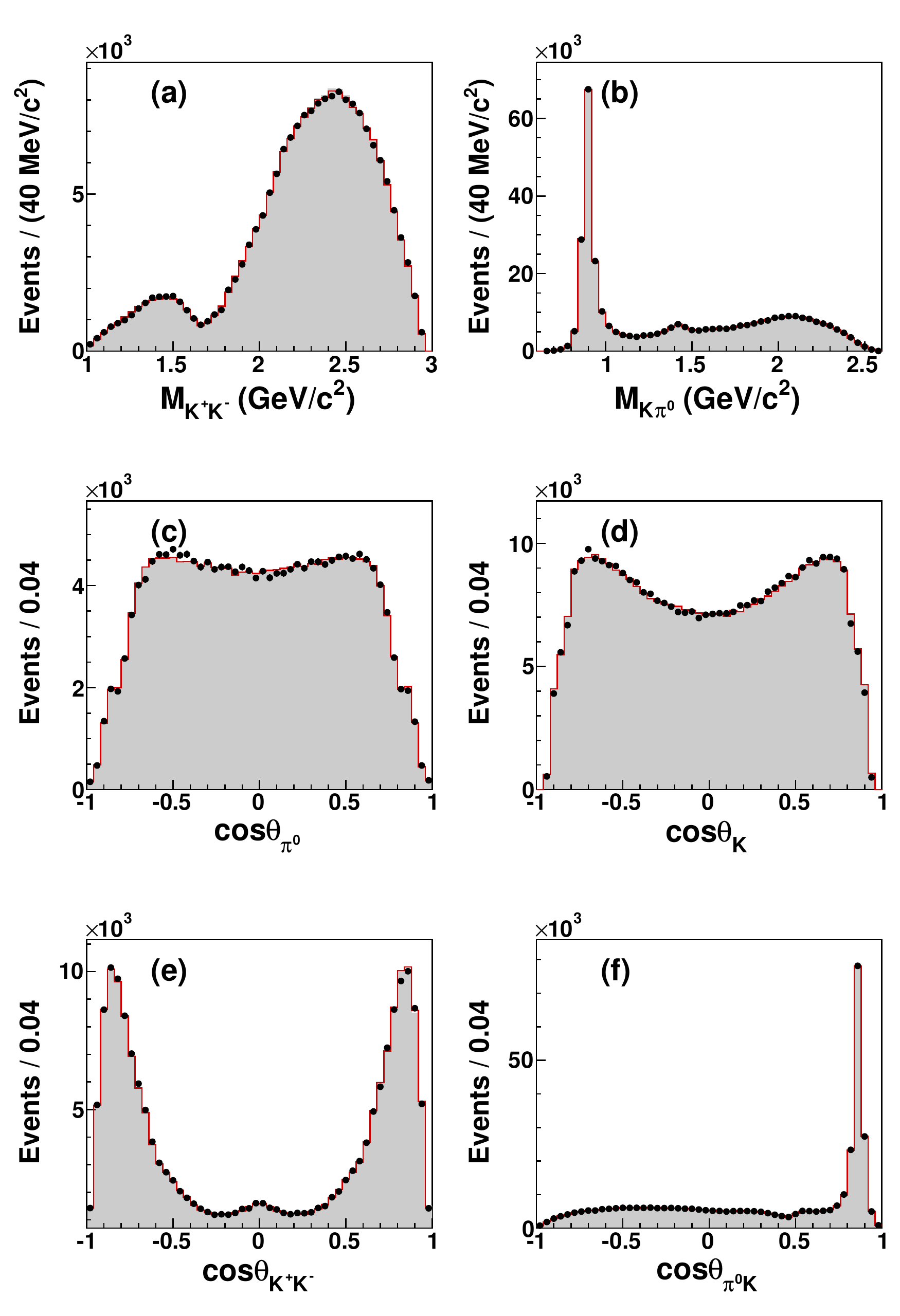}}
\caption{Kinematical distributions for the data (dots),
the PWA solution~I (shaded histograms) and the PWA solution~II (solid line).
The notation $K$ without any specified charge indicates the sum of the
$K^+$ and $K^-$ distributions.
(a-b) Invariant mass of the $K^+K^-$ and $K^\pm\pi^0$ systems.
(c-d) Distributions of the final state particles polar angle ($\theta_{\pi^0}$, $\theta_{K}$)
with respect to the beam axis in the $J/\psi$ rest frame.
(e-f) Polar angle distributions ($\theta_{KK}$, $\theta_{\pi K}$) for $K^+$ in the $K^+K^-$ 
helicity frame (e) and for $\pi^0$ in the $K\pi^0$
helicity frame (f). The uncertainties are statistical and are within the size of the dots.
}
\label{fig:kin_distributions_full}
\end{figure}

\begin{figure}[t]
\centerline{\includegraphics[width=0.48\textwidth]{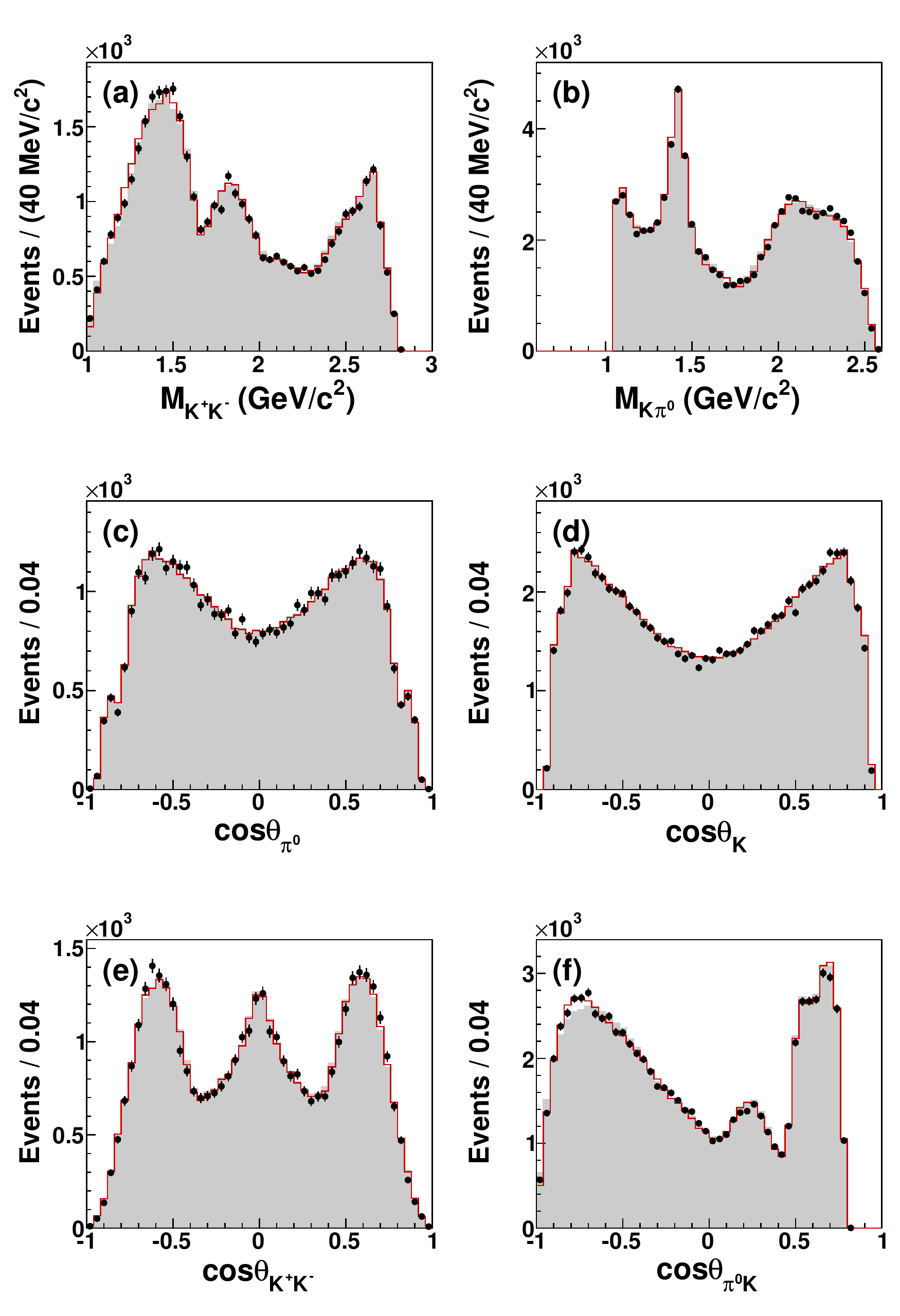}}
\caption{Kinematical distributions for the data (dots),
PWA solution I (shaded histograms) and PWA solution II (solid line)
in the inner region of the Dalitz
plot ($M(K^\pm\pi^0)>1.05$ GeV/$c^2$).
The notation $K$ without any specified charge indicates the sum of the
$K^+$ and $K^-$ distributions.
(a-b) Invariant mass of the $K^+K^-$ and $K^\pm\pi^0$ systems.
(c-d) Distributions of the final-state state particles polar angle ($\theta_{\pi^0}$, $\theta_{K}$)
with respect to the beam axis in the $J/\psi$ rest frame.
(e-f) Polar angle distributions ($\theta_{KK}$, $\theta_{\pi K}$) for $K^+$ in the $K^+K^-$ 
helicity frame (e) and for the $\pi^0$ in the $K\pi^0$
helicity frame (f).
The  error bars represent the statistical uncertainties. 
}
\label{fig:kin_distributions}
\end{figure}

The results for the best fit based on the well-established
resonances and constant contributions in the lowest partial waves
are given in Table~\ref{tab:fbestPDG}.
Only contributions improving the NLL by more than 17
are included to the fit (corresponding to a statistical significance
of $5\sigma$ for 4 degrees of freedom).
The data description as a Dalitz plot are shown in Fig.~\ref{fig:dalitz}(b).
Fig.~\ref{fig:kin_distributions_full}
and Fig.~\ref{fig:kin_distributions} show the corresponding
invariant mass spectra and angular distributions.
The kinematic distributions in Fig.~\ref{fig:kin_distributions}
are restricted to the inner part of the Dalitz plot
($M(K^\pm\pi^0)>1.05$~GeV/$c^2$) to exclude the huge
peaks from the $K^*(892)^\pm$.

The dominant contribution stems from the $K^*(892)^\pm$ and $K_2^*(1430)^\pm$ 
resonances in the $K^\pm\pi^0$ kinematic channels.
The first decay is well-known and contributes about 90\% to the
total decay rate. The interference term between the $K^{*}(892)^+K^-$
and  $K^{*}(892)^-K^+$ intermediate states contributes about 10\%.
The mass and the width of the $K^*(892)^\pm$ are determined with high
statistical precision.
The Blatt-Weisskopf radius of the resonance is found to be
$r=0.25\pm0.02$~fm. The second largest contribution,
with a decay fraction of about 10\%, is the 
$K^*_2(1430)^\pm$, which also can be clearly seen
in Fig.~\ref{fig:dalitz}.
The mass and width of this state are also determined with
high precision. Its Blatt-Weisskopf radius can not be
reliably determined from the fit and is set
to $0.4$~fm, which is the meson-interaction radius
used in Ref.~\cite{Aston:1986rm}.
The contribution of the
$K^*_2(1430)^\pm K^\mp$ channel to the reaction is approximately 10
times smaller than the contribution from the $K^*(892)^\pm K^\mp$
channel. Taking into account this result and
using a branching fraction of 49.9\% for the $K_2^*(1430)^\pm$ decay to
$K\pi$ \cite{Tanabashi:2018oca}, we find that the 
$J/\psi$ decay to $K^*_2(1430)^\pm K^\mp$ is suppressed
by an approximate factor of 5 compared to the decay to
$K^*(892)^\pm K^\mp$.
For $J^P=1^-$, the inclusion of the $K^*(1680)^\pm$
provides a significant improvement in the data description,
but no NLL minima consistent with its mass and width are found.
The $J^P=2^+$ partial wave requires another $2^+$ state
with a relative contribution of approximately 0.4\%.
Its mass and width are found to be $1817\pm11$ MeV/$c^2$ and $312\pm28$ MeV/$c^2$,
respectively. This mass is much lower than the mass of the $K_2^*(1980)^\pm$
observed by LASS. The $K_3^*(1780)^\pm$ state
provides a significant improvement in the log-likelihood, but
no NLL minima consistent with its measured parameters
are found. Finally, there is a small , but
very distinct and stable
contribution of $(0.18\pm0.02)$\% from the $K_4^*(2045)^\pm$. Its fitted mass is lower
than that obtained in other measurements~\cite{Tanabashi:2018oca}, which
can be attributed to the uncertainties of the PWA solution
(see solution~II).

In the $K^+K^-$ kinematic channel, the first stable contribution has
$J^{PC}=1^{--}$, a mass of $1643\pm3$~MeV/$c^2$,
a width of $167\pm12$~MeV/$c^2$ and a decay fraction of 1\%.
It can also be clearly seen in the
Dalitz plot. As mentioned above, this contribution can be attributed to the 
$\rho(1700)$. The structure is also reasonably consistent
with the $\omega(1650)$ (the mass is consistent with the PDG estimate,
and the width is well within the spread of the results quoted by PDG)
or an interference between these states.
The second contribution that can be reliably determined from the data
is a $J^{PC}=1^{--}$ resonance with a mass of $2078\pm6$~MeV/$c^2$ and
width of $149\pm21$ MeV/$c^2$. The largest relative contribution of $(1.8\pm0.2)$\%
comes from the tail of the $\rho(770)$. Since the mass of this state is significantly below
the $K^+K^-$ production threshold,
no reliable claim can be made about its observation.
The $\rho_3(1690)$ and $\rho(1450)$ provide NLL
improvement by 144 and 27, but no NLL
minimum consistent with the parameters of each state is found.
The smooth contribution in the $J^{PC}=1^{--}$ $K^+K^-$
partial wave is also found to be significant.

Additionally, we try to set the mass and the width
of the $J^{PC}=1^{--}$ $K^+K^-$ contribution at
1.65~GeV/$c^2$ to the PDG mean values for the $\rho(1700)$
averaged from $\eta\rho(770)$ and $\pi^+\pi^-$ modes.
In this case, the NLL worsens by 42, and so one
may consider  including the  $\omega(1420)$ and $\omega(1650)$ in
the fit. In these fits we set their masses and width
to the mean values of the PDG estimates. If the $\omega(1420)$ ($\omega(1650)$)
is included, the NLL is still worse by 14 (7) compared
to the result of solution~I.
If the $\rho(1450)$ is substituted by the $X(1575)$, instead of
adding a resonance, the NLL improves by 28, but
remains worse by 14 than the result for solution~I.

Adding further well-established resonances with the nominal 
PDG parameters does not improve log-likelihood by more than 17 units.
Despite this, the solution is not saturated: if additional
contributions (parametrized as Breit-Wigner resonances with
parameters not required to correspond to a physical state)
are added, they can improve NLL
by up to 95 in a single partial wave, which is much
larger than the contribution of other resonances included
to the solution. The only notable additional contribution
indicating resonance behavior is in the $J^P=1^-$ $K\pi$
partial wave with a mass of around 2.4 GeV/$c^2$, but there
is lack of qualitative evidence to report a new state.
The largest improvement
in the NLL function comes from contributions
that tend to be broad and cannot be interpreted as resonances.
These conclusions are not surprising if one considers the measured two-particle
$K\pi$ scattering amplitudes obtained by the LASS Collaboration \cite{Aston:1987ir}.
Here the $F$-wave intensity, apart from the $K^*_3(1780)$ peak,
has a strong contribution from nontrivial structures,
which are not resolved in the LASS analysis.
The inability to provide a consistent data description
for this solution prevents us from making a reliable 
estimation of systematic uncertainties.

\begin{table*}[htb]
\footnotesize
\renewcommand{\arraystretch}{1.6}
  \caption{
List of contributions for solution~I, showing for each contribution the mass, width, decay fraction and increase in negative log-likelihood for the removal of the state.
In the $K\pi$ channel $b$ stands for decay fraction through
both charged conjugated modes and $b^{+(-)}$ gives the contribution
of one charged mode, which allows their interference to be determined.
The uncertainties are statistical. Parameters marked with $^\star$ are fixed.
}
  \label{tab:fbestPDG}
  \center
  \setlength{\tabcolsep}{11pt}
  \begin{tabular}{  >{$}l<{$}  >{$}l<{$}  >{$}r<{$}  >{$}r<{$}  >{$}r<{$}  >{$}r<{$}  >{$}r<{$}  } \hline\hline
   \multicolumn{7}{ c }{\textbf{$K^\pm\pi^0$ channels}} \\
   J^{PC}\!\!      & \text{PDG}        & M(\text{MeV/}c^2) &\Gamma (\text{MeV/}c^2)  & b(\%)  & b^{+(-)}(\%) & \Delta\text{NLL} \\     
   \hline\hline
      \;1^{-}\!\!     & K^*(892)^\pm     & 894.1 \mpm0.1                  & 46.7 \mpm0.2                   & 89.2\mpm0.8                  & 41.0\mpm0.2                    &      -    \\ \hline
      \;1^{-}\!\!     & K^*(1680)^\pm    & 1677^{\star }                      & 205^{\star }                       & 0.59\mpm0.04                 & 0.25\mpm0.02                   &     398   \\ \hline
      \;2^{+}\!\!     & K_2^*(1430)^\pm  & 1431.4\mpm0.8                  & 100.3\mpm1.6                   & 9.2 \mpm0.1                  & 4.1 \mpm0.1                    &       -   \\ \hline
      \;2^{+}\!\!     & K_2^*(1980)^\pm  & 1817\mpm11                     & 312\mpm28                      & 0.44\mpm0.05                 & 0.17\mpm0.02                   &     238   \\ \hline
      \;3^{-}\!\!     & K_3^*(1780)^\pm  & 1781^{\star }                      & 203^{\star }                       & 0.08\mpm0.01                 & 0.04\mpm0.01                   &      83   \\ \hline
      \;4^{+}\!\!     & K_4^*(2045)^\pm  & 2015\mpm7                      & 183\mpm17                      & 0.16\mpm0.02                 & 0.07\mpm0.01                   &     192   \\ \hline
\multicolumn{7}{ c }{\textbf{$K^+K^-$ channel}} \\
\quad\;\, J^{PC}\!\!      & \text{PDG}        & M(\text{MeV/c}^2) &\Gamma (\text{MeV/c}^2)  & \multicolumn{2}{ c }{b(\%)}  & \Delta\ln L \\ \hline
   \quad\;\,   1^{--}\!\! &\rho(770)           & 771^{\star}                  & 150^\star                          & \multicolumn{2}{ c }{$1.8 \mpm0.2                 $} &     220  \\ \hline
   \quad\;\,   1^{--}\!\! &\rho(1450)          & 1465^\star                   & 400^\star                          & \multicolumn{2}{ c }{$1.2 \mpm0.2                 $} &      27  \\ \hline
   \quad\;\,   1^{--}\!\! &                    & 1643  \mpm3              & 167  \mpm12                    & \multicolumn{2}{ c }{$1.1 \mpm0.1                 $} &     281  \\ \hline
   \quad\;\,   1^{--}\!\! &                    & 2078\mpm6                & 149\mpm21                      & \multicolumn{2}{ c }{$0.15\mpm0.03                $} &      73  \\ \hline
   \quad\;\,   1^{--}\!\! & \text{non-resonant}      &      --                        &    --                        & \multicolumn{2}{ c }{$1.2 \mpm0.2       $} &      34  \\ \hline
   \quad\;\,   3^{--}\!\! &\rho_3(1690)        & 1696^\star                   & 204^\star                          & \multicolumn{2}{ c }{$0.14\mpm0.01                $} &     144  \\
   \hline\hline
  \end{tabular}
\end{table*}

\subsection{Solution II}
We find that the largest improvement to the NLL
of the solution~I comes from the inclusion of a smooth contribution
in the $J^P=3^-$ partial wave, which we parametrize with a broad Breit-Wigner shape.
Its mass is found to be close the maximal allowed
invariant mass of the $K^\pm\pi^0$ system.
The width can vary in the approximate interval of  0.5~GeV/$c^2$~--~1.2~GeV/$c^2$,
depending on small variations of the PWA solution, and its
value only slightly affects other components in the fit.
Such a mass and width does not allow an interpretation of
this contribution as a single resonance.
The solution where this broad component is added
and the significance of the resonances is reevaluated is shown in Table~\ref{tab:fbestres}.
For this solution, we use the more conservative resonance
significance criteria: the minimum NLL improvement is
required to be 40. 
We ensure that no other allowed resonance
contributions improve the NLL value above
this number, considering possibilities with spins up
to $J=5$, which is the maximum  spin of previously reported
states allowed in this decay.
Those contributions which give the most significant
NLL improvement are used to estimate systematic
uncertainties. The NLL value for
this solution is better by 116 than that of solution~I.
The systematic uncertainties listed in Table~\ref{tab:fbestres}
will be discussed later. The Dalitz plot for the solution~II
is shown in Fig.~\ref{fig:dalitz}(c).  
Mass and angular distributions are given in Fig.~\ref{fig:kin_distributions_full}
and Fig.~\ref{fig:kin_distributions} for the data and for the two models. 
The two descriptions are very similar, but solution~II is superior in specific kinematic regions.

Solution~II has the same set of  well-defined contributions
as solution~I. The fitted mass and
width for the $K^*(892)^\pm$ and $K_2^*(1430)^\pm$ are
almost the same. The mass, width and Blatt-Weisskopf radius
of the $K^*(892)$ are found to be
$M=893.6 \mpm0.1 _{-0.3 }^{+0.2}$~MeV/$c^2$, 
$\Gamma=46.7 \mpm0.2 _{-0.2 }^{+0.1 }$~MeV/$c^2$ and
$r=0.20\pm0.02_{-0.04}^{+0.14}$~fm, respectively, where here
and subsequently the first uncertainty is statistical, and the second systematic.
The mass lies between
the PDG averages for measurements performed 
where the $K^*(892)^\pm$ is produced
in hadronic collisions and those were it is produced in $\tau$ decays~\cite{Tanabashi:2018oca}.
The fitted width is consistent with the $\tau$-decay results~\cite{Epifanov:2007rf}. 
For the $K_2^*(1430)^\pm$ we fix
the Blatt-Weisskopf radius to 0.4~fm. The $2^+$ partial amplitude
in the $K^\pm\pi^0$ kinematic channels also requires a second
contribution with a mass higher than that of the previous solution
with large systematic uncertainties for both the mass and width: $M=1868\pm8_{-57}^{+40}$ MeV/$c^2$
and $\Gamma=272\pm24_{-15}^{+50}$ MeV/$c^2$. The mass is approximately
100~MeV/$c^2$ below the LASS measurement for the $K_2^*(1980)$ \cite{Aston:1986jb}, but
both the mass and the width are compatible with the PDG averages within $2.2$
standard deviations.
As in solution~I, there is a very clear contribution to the $J^P=4^+$ partial wave
with $M=2090\pm9^{+11}_{-29}$~MeV/$c^2$ and $\Gamma=201\pm19_{-17}^{+57}$~MeV/$c^2$,
which is consistent with the parameters of the $K_4^*(2045)^\pm$ \cite{Tanabashi:2018oca}.
For the $K^*(1410)$, which is required in this solution, the
$K^*(1680)^\pm$ and the $K_3^*(1780)^\pm$, no NLL minima
consistent with parameters of these resonances are found. In the
$K^+K^-$ kinematic channel we see again two
stable contributions at 1.65~GeV/$c^2$ and 2.05 GeV/$c^2$. The contributions
from the $\rho(1450)$, $\rho_3(1690)$ and $\rho(770)$ are marginal.

A striking feature of solution~II is the presence
of a non-resonance component in the  $J^P=3^-$ $K^\pm\pi^0$ partial
waves, which can not be clearly interpreted as an
interference between Breit-Wigner states. A possible interpretation is that this component
is the manifestation of  non-resolved contributions present in the
$F$-wave $K\pi$ scattering amplitude \cite{Aston:1987ir}.
This may include the presence of several resonances, non-resonant production and
final-state particle rescattering effects.

\begin{figure*}
\centerline{
      \includegraphics[width=0.95\textwidth]{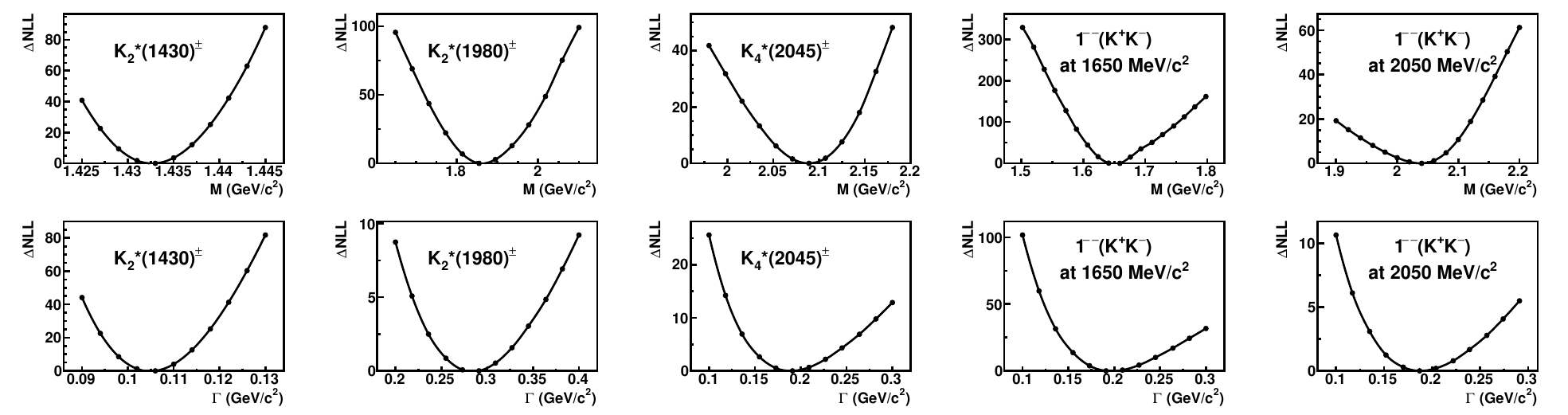}
  }
\caption{\label{fig:mass_scan} Mass and width scans for the $K^*_2(1430)$,
$K_2^*(1980)$, $K^*_4(2045)$ and $1^{--}$ structures at 1650 MeV/$c^2$
and 2050 MeV/$c^2$ for solution~II.}
\end{figure*}

The stability of the found NLL minimum with respect to
the parameters of the reported resonances is demonstrated
in Fig.~\ref{fig:mass_scan}.

The systematic errors due to the uncertainty of the PWA solution
are assigned to be the largest deviations for the following
variations of the solution:
\begin{itemize}
    \item variation of the masses and widths for the $K^\pm\pi^0$ resonances with
        the parameters fixed in the fit, and varied by one standard
        deviation of the LASS results \cite{Aston:1987ir};
    \item variation of the Blatt-Weisskopf radius of the $K_2^*(1430)^\pm$ by $\pm0.2$~fm;
    \item inclusion of contributions that strongly improve
        the log-likelihood below the acceptance criteria
        ($J^P=1^{-}$ $(K\pi)$ at approximately 2.5 GeV/$c^2$
        and $J^{PC}=1^{--}$ $(K^+K^-)$ at
        $M(K^+K^-)\approx 2.3$ GeV/$c^2$);
    \item reparametrization of the broad background part of partial waves.
\end{itemize}
To evaluate the latter variation, broad contributions in the $1^-$, $2^+$ ($K\pi$)
amplitudes and $1^{--}$ ($K^+K^-$) partial wave parametrized with
$\rho(770)^0$ and $\rho(1450)^0$ are studied.  In all these fits the states
$K^*(892)^\pm$, $K_2^*(1430)^\pm$, $K_4^*(2045)^\pm$ and the structures at
1.65~GeV/$c^2$ and 2.05~GeV/$c^2$ in the $K^+K^-$ channels remain stable.
The high-mass 
broad $K^\pm\pi^0$ $3^-$ contribution always remains significant, but its relative
fraction varies to much smaller values in some fits.
The $1^-$ additional contribution mostly manifests
resonant behavior. No stable contribution can be associated with the 
$\rho(1450)$, but its relative decay fraction at the level of 1\%
does not contradict the data.

The total systematic uncertainties for the masses, widths and decay fraction given in Table~\ref{tab:fbestres} are calculated as a quadratic sum of:
\begin{itemize}
    \item the variation in results due to the uncertainty of the PWA solution;
    \item the bias introduced by imperfections of the detector simulation and
        the event reconstruction;
    \item the uncertainties  due to the differences in kaon tracking and PID
    efficiencies between data and the MC simulation.     
\end{itemize}
The differences
in kaon tracking and PID
efficiencies between data and the MC simulation
are studied with a high-purity
control sample of $J/\psi \to K_SK^\pm\pi^\mp$ decays as
a function of kaon transverse momentum $p_T$ and are
found to be within $1\%$ per track both for the tracking
and the PID. The effect on the
PWA result is estimated by varying the selection efficiency
difference for data and MC in $p_T$ bins within
these errors. Uncertainties on the fit parameters due to the efficiency
variation in each bin are summed quadratically.

The background uncertainty, estimated by varying the
subtracted NLL contribution
by 50\%, is found to be negligible.

\begin{table*}[htb]
\footnotesize
\renewcommand{\arraystretch}{1.6}
  \caption{
List of components for solution~II.
For the reported states
in the $K\pi$ channel  ($K^*(892)^\pm$, $K_2^*(1430)^\pm$, $K_2^*(1980)^\pm$ and $K_4^*(2045)^\pm$) and the reported signals in the $K^+K^-$ channel ($J^{PC}=1^{--}$
signals with masses around 1650 MeV/$c^2$ and 2050 MeV/$c^2$) the first uncertainty is statistical and the second is systematic.
In the $K\pi$ channel the  decay fraction is given for both charged conjugated
modes ($b$) and for the contribution of one charged mode  ($b^{+(-)}$), so that 
their interference can be determined.
As the $K^*(1410)^\pm$, $K^*(1680)^\pm$ and $K_3^*(1780)^\pm$ contributions are not
reliably identified (see main text), their masses and widths are fixed (marked with $^\star$)
and only statistical uncertainties are given  for their decay fractions.
}
  \label{tab:fbestres}
  \center
  \setlength{\tabcolsep}{11pt}
  \begin{tabular}{  >{$}l<{$}  >{$}l<{$}  >{$}r<{$}  >{$}r<{$}  >{$}r<{$}  >{$}r<{$}  >{$}r<{$}  } \hline\hline
   \multicolumn{7}{ c }{\textbf{$K^\pm\pi^0$ channels}} \\
   J^{PC}\!\!      & \text{PDG}        & M(\text{MeV/}c^2) &\Gamma (\text{MeV/}c^2)  & b(\%)  & b^{+(-)}(\%) & \Delta\text{NLL} \\     
   \hline\hline
      \;1^{-}\!\!     & K^*(892)^\pm   & 893.6 \mpm0.1 _{-0.3 }^{+0.2 } & 46.7 \mpm0.2 _{-0.2 }^{+0.1 }  & 93.4\mpm0.4_{-5.8 }^{+1.8 }    & 42.5\mpm0.1_{-1.7 }^{+0.5 }      &    -      \\ \hline
      \;1^{-}\!\!     & K^*(1410)^\pm  & 1380^{\star }                      & 176^{\star }                       & 0.26\mpm0.04                 & 0.11\mpm0.02                   &      80   \\ \hline
      \;1^{-}\!\!     & K^*(1680)^\pm  & 1677^{\star }                      & 205^{\star }                       & 0.20\mpm0.03                 & 0.08\mpm0.01                   &      56   \\ \hline
      \;2^{+}\!\!     & K_2^*(1430)^\pm  & 1432.7\mpm0.7_{-2.3}^{+2.2}    & 102.5\mpm1.6_{-2.8}^{+3.1}      & 9.4 \mpm0.1 _{-0.5 }^{+0.8 } & 4.2 \mpm0.1 _{-0.2 }^{+0.3 }   &       -   \\ \hline
      \;2^{+}\!\!     & K_2^*(1980)^\pm  & 1868\mpm8_{-57}^{+40}          & 272\mpm24_{-15}^{+50}          & 0.38\mpm0.04_{-0.05}^{+0.22} & 0.15\mpm0.02_{-0.02}^{+0.08}   &     192   \\ \hline
      \;3^{-}\!\!     & K_3^*(1780)^\pm  & 1781^{\star }                      & 203^{\star }                       & 0.16\mpm0.02                 & 0.07\mpm0.01                   &     105   \\ \hline
      \;4^{+}\!\!     & K_4^*(2045)^\pm  & 2090\mpm9_{-29}^{+11}          & 201\mpm19_{-17}^{+57}          & 0.21\mpm0.02_{-0.05}^{+0.10} & 0.09\mpm0.01_{-0.02}^{+0.04}   &     212   \\ \hline
   \;3^{-}\!\!      & \text{non-resonant}       & --   & --  & \sim 1.5\%     & \sim0.6\%     &     629   \\
\hline
\multicolumn{7}{ c }{\textbf{$K^+K^-$ channel}} \\
\quad\;\, J^{PC}\!\!      & \text{PDG}  & M(\text{MeV/c}^2)    &\Gamma (\text{MeV/c}^2)      & \multicolumn{2}{ c }{b(\%)}                          & \Delta\ln L \\ \hline
   \quad\;\,   1^{--}\!\! &      & 1651\mpm3_{-6}^{+16}        & 194\mpm8_{ -7}^{+15}        & \multicolumn{2}{ c }{$1.83\mpm0.11_{-0.17}^{+0.19}$} &     796  \\ \hline
   \quad\;\,   1^{--}\!\! &      & 2039\mpm8  _{-18}^{+36}     & 193\mpm23_{-27}^{+25}       & \multicolumn{2}{ c }{$0.23\mpm0.04_{-0.06}^{+0.07}$} &     102  \\
   \hline\hline
  \end{tabular}
\end{table*}

\subsection{Summary on PWA}
Our analysis shows that there is a set of states in the PWA
solutions that remains stable for both considered cases:
when contributions corresponding to well-known
resonances are considered or when broad contributions are introduced
to parameterize the missing part of the partial amplitudes.
In the $K^\pm\pi^0$ channels this set of resonances includes
the $K^*(892)^\pm$, $K_2^*(1430)^\pm$, and $K_4^*(2045)^\pm$.
The second $J^P=2^+$ state, labeled here as $K_2^*(1980)^\pm$, has a mass
much lower than that observed by the LASS Collaboration \cite{Aston:1986jb}.
However, given the large systematic uncertainties on
this quantity, our result  is compatible within 2.2~standard deviations.
The first stable structure in the $K^+K^-$ channel has a mass of about 1.65~GeV/$c^2$
and a decay fraction of 1.0\%~--~1.5\%. The absence of a distinct
contribution from the first radial excitation of the $\rho(770)$ favors its interpretation
as a $^3D_1$ $\rho$-resonance. At the same time such a small decay fraction is
consistent with $\omega(1650)$ production in $J/\psi$ decay through
a virtual photon.
Its mass is consistent with the PDG estimate for the $\omega(1650)$ and its
width is well within the spread of experimental results quoted by the PDG.
It could also be the result of interference between these isovector and isoscalar states.
The second stable contribution has a mass of about 2.05~GeV/$c^2$~--~2.10~GeV/$c^2$ and
decay fraction of 0.1\%~--~0.2\%. Given the large systematic uncertainties
it could be interpreted as either the $\rho(2150)$ or as  another
isovector-vector state observed in 
proton-antiproton annihilation in flight \cite{Anisovich:2000ut}.
Clarification of the nature of these excited vector mesons requires
further investigation.

\section{Branching fractions}

The $J/\psi \to K^+K^-\pi^0$ branching fraction is determined as
$B(J/\psi \to K^+K^-\pi^0) =
\frac{N_{sel}-N_{bg}-N_{continuum}}{\epsilon N_{J/\psi} B(\pi^0 \to
\gamma\gamma)}$. Here $N_{sel}$, $N_{bg}$ and $N_{continuum}$ are
the number of selected events, the estimated background from the $J/\psi$
decays, and the continuum production, respectively. The number of $J/\psi$ events
$N_{J/\psi} = (223.7\pm 1.4(syst.)) \times 10^6$ is taken from Ref.~\cite{Ablikim:2016fal}, and $B(\pi^0 \to \gamma\gamma) = (98.823 \pm 0.034) \times 10^{-2}$
is taken from the PDG~\cite{Tanabashi:2018oca}. The selection efficiency $\epsilon$
is obtained using the PWA solution~II and the detector performance simulation.
The dominant contribution to the statistical uncertainty comes from 
 $N_{sel}$.
The systematic uncertainty on the branching fraction is estimated
from the sources listed in Table~\ref{tab:brsyst_summ}.
The background uncertainty is estimated by
varying $N_{bg}$ by $\pm50\%$.
The uncertainty associated with the subtraction of the continuum background is assigned to be 
the statistical error on 
$N_{continuum}$.
The charged track reconstruction
efficiency and the PID efficiency uncertainties are 1\% each per
track as  is discussed above. The photon detection efficiency is
studied with the decays $\psi(3686) \to \pi^+\pi^- J/\psi$, $J/\psi \to
\rho^0\pi^0$ and photon conversion control
samples \cite{Ablikim:2010zn, Ablikim:2011kv}. In this analysis,
an uncertainty of 1\% per photon is assigned.
The uncertainty introduced by the cut on
$\chi^2_{K^+K^-\gamma\gamma}$ is estimated using a control sample.
This is selected using similar selection criteria,
with the kinematic-fit cut replaced
by the requirement that at least one particle out of three
($K^+$, $K^-$, $\pi^0$) has a mass hypothesis consistent
with the recoil mass calculated using the other two particles.
Such a procedure accepts a signal event even if one of the
particles is badly reconstructed.
This gives
$B(J/\psi \to K^+K^-\pi^0)=(2.88\pm0.01\pm0.12)\times10^{-3}$.

\begin{table}[h]
\centering
\caption{Summary of systematic uncertainties for $B(J/\psi \to K^+K^-\pi^0)$.}
\label{tab:brsyst_summ}
\begin{tabular}{ l c }
       \hline\hline
              Source                   & Uncertainty (\%) \\ \hline\hline
$N_{bg}$                               &        0.2         \\
$N_{continuum}$                        &        0.3         \\
Track reconstruction efficiency        &        2.0         \\
PID efficiency                         &        2.0         \\
Photon reconstruction efficiency       &        2.0         \\
Kinematic fit cut efficiency           &        2.4         \\
$N_{J/\psi}$\cite{Ablikim:2016fal}     &        0.6         \\
\hline
Total                                  &        4.3         \\
\hline\hline
\end{tabular}
\end{table}

Knowing the $J/\psi \to K^+K^-\pi^0$ branching fraction and the
decay fractions for the individual components from the PWA, we determine branching
fractions for the decay via individual resonances. Results
for solution~II are summarized in Table~\ref{tab:resonance_br}.
The branching fraction $B(J/\psi \to K^+K^-\pi^0)$ and the branching fractions
for the decay via the $K^*(892)^\pm$ that are obtained in solution~II
are compared to the results
from previous experiments in Table~\ref{tab:br_compar}.
Our result for $B(J/\psi \to K^+K^-\pi^0)$
is up to now the most precise measurement. It differs
from the PDG value~\cite{Tanabashi:2018oca}, obtained indirectly from Ref.~\cite{Aubert:2007ym},
by about 2.8 standard deviation.
The systematic uncertainty of our results for decays through the $K^*(892)^\pm$
is somewhat larger than that of Ref.~\cite{Aubert:2007ym},
which can be attributed to  the uncertainties present in the PWA model.

\begin{table*}[h]
    \small
\renewcommand{\arraystretch}{1.6}
    \centering
    \caption{Branching fractions for decays via reliably identified intermediate states (solution~II). 
    $R_{K\pi}$ and $R_{KK}$ denotes $K^\pm\pi^0$ and $K^+K^-$ resonances, respectively, and
    $R_{K\pi}^\pm K^\mp$ denotes for one possible charged combination. The first uncertainty
    is statistical and the second one is systematic.}
    \label{tab:resonance_br}
    \begin{tabular}{ >{$}l<{$}  >{$}r<{$}  >{$}r<{$} }
        \hline\hline
        \multicolumn{3}{ c }{\textbf{Intermediate resonance in the $K\pi$ system}} \\ \hline
        \text{$R_{K\pi}$}   & B(J/\psi \to R_{K\pi}^\pm K^\mp \to K^+K^-\pi^0)  & B(J/\psi \to R_{K\pi}^+ K^-  + c.c.  \to K^+K^-\pi^0)   \\
        \hline
        K^*(892)         &  (1.22\mpm0.01  \er{0.07}{0.05})  \times10^{-3}  &  (2.69\mpm0.01  \er{0.20}{0.13})  \times10^{-3}    \\ \hline
        K_2^*(1430)      &  (1.21\mpm0.02  \er{0.08}{0.10})  \times	10^{-4}  &  (2.69\mpm0.04  \er{0.19}{0.25})  \times10^{-4}    \\ \hline
        K_2^*(1980)      &  (4.3\mpm0.5    \er{0.6}{2.3})    \times10^{-6}  &  (1.1\mpm0.1    \er{0.1}{0.6})    \times10^{-5}    \\ \hline
        K_4^*(2045)      &  (2.6\mpm0.3    \er{0.6}{1.1})    \times10^{-6}  &  (6.2\mpm0.7    \er{1.4}{2.8})    \times10^{-6}    \\ \hline
        \multicolumn{3}{ c }{\textbf{Intermediate resonance in the $K^+K^-$ system}} \\ \hline
        \text{$R_{KK}$}   & \multicolumn{2}{ c }{$B(J/\psi \to R_{KK}\pi^0  \to K^+K^-\pi^0)$}  \\\hline
        1^{--}(1650\text{ MeV/}c^2)   &   \multicolumn{2}{ c }{$(5.3\mpm0.3    \er{0.5}{0.6})  \times10^{-5} $}   \\ \hline
        1^{--}(2050\text{ MeV/}c^2)   &   \multicolumn{2}{ c }{$(6.7\mpm1.1    \er{1.8}{2.2})    \times10^{-6} $}   \\
        \hline\hline
    \end{tabular}
\end{table*}

\begin{table*}[h]
\renewcommand{\arraystretch}{1.6}
\centering
\scriptsize
\caption{Comparison between this work and previous measurements.
	For  $B(J/\psi\to K^{*+}K^-\!+\!c.c. \to K^+K^-\pi^0)$ and
	$B(J/\psi\to K^{*+}K^-\!+\!c.c.)$ we give two numbers for
solution~II: the first one
is a sum of branching fractions through $K^{*+}$ and $K^{*-}$ and
the second number (in parenthesis) accounts for their interference.
Results marked with ``$^\dagger$'' are obtained by averaging the $K_SK^\pm\pi^\mp$ and
$K^+K^-\pi^0$ final states. Results recalculated by us using numbers from this work
are marked with ``$^{\dagger\dagger}$''.}
\begin{tabular}{ p{.27\textwidth}   r   r   r   r   r   r}
\hline\hline
Channel                                               &\multicolumn{5}{c }{$B(\times10^{-3})$}                                        \\ \cline{2-6}
                            &This work           &BABAR\cite{Aubert:2007ym} & DM2\cite{Jousset:1988ni} & MARK-III\cite{Coffman:1988ve} & MARK-II\cite{Franklin:1983ve} \\ \hline
$B(J/\psi\to K^+K^-\pi^0)$ &$2.88\!\pm\!0.01\!\pm\!0.12   $ & ---                      &  ---                     &  ---                      & $2.8\!\pm\!0.8$ \\ \hline
$B(J/\psi\to K^{*+}K^-\!+\!c.c.\!\to\!K^+K^-\pi^0)$  &$2.45\!\pm\!0.01_{-0.14}^{+0.10}(2.69\!\pm\!0.01_{-0.20}^{+0.13}) $ &$1.97\!\pm0.16\!\pm0.13$    & $1.50\!\pm\!0.23\!\pm\!0.27^{\dagger\dagger}$ & $1.87\!\pm\!0.04\!\pm\!0.28^{\dagger\dagger}$             & $2.6\!\pm\!0.8$         \\ \hline
$B(J/\psi\to K^{*+}K^-\!+\!c.c.)$                    &$7.34\!\pm\!0.03_{-0.43}^{+0.33}(8.07\!\pm\!0.04_{-0.61}^{+0.38}) $ &$5.2\!\pm\!0.3\!\pm\!0.2^\dagger$ & $4.57\!\pm\!0.17\!\pm\!0.70^\dagger$    & $5.26\!\pm\!0.13\!\pm\!0.53^\dagger$    & $7.8\!\pm\!2.4^{\dagger\dagger}$ \\
\hline\hline
\end{tabular}
\label{tab:br_compar}
\end{table*}

\section{Conclusion}

A partial-wave analysis of the decay $J/\psi \to K^+K^-\pi^0$
using a data sample of $(223.7\pm1.4)\times10^6$ $J/\psi$
events collected by the BESIII reveals a set of resonances that
have not been observed by previous experiments.
In the $K^\pm\pi^0$ channels our analysis reveals signals from
$K_2^*(1980)^\pm$ and $K_4^*(2045)^\pm$ resonances. This is the first
observation of these states in $J/\psi$ decays. The mass
of the former state is determined with 
a central value around 100 MeV/$c^2$ lower than that
reported by the LASS Collaboration~\cite{Aston:1986rm}. This lower value is in
better agreement with the expectation from the linear Regge trajectory of
radial excitations with the standard slope \cite{Anisovich:2001ig}.
As for the known decays through $K\pi$ resonances, we
determine the parameters, decay ratios, and branching fractions
for the $K^*(892)^\pm$ and $K_2^*(1430)^\pm$ with improved precision
compared to previous measurements.
In the $K^+K^-$ channel we observe a clear $J^{PC}=1^{--}$ resonance structure with a
mass of 1.65~GeV/$c^2$ and another $J^{PC}=1^{--}$ contribution at 2.05~GeV/$c^2$~--~2.10~GeV/$c^2$. The
first structure may be interpreted as the ground $^3D_1$ isovector state.
At the same time its mass, width and small relative contribution
to the decay are reasonably consistent
with the production of the $\omega(1650)$
in $J/\psi$ decays through a virtual photon.
The second state can be interpreted as the $\rho(2150)$ or as another
isovector-vector state that has been observed in
proton-antiproton annihilation in flight \cite{Anisovich:2000ut}.
The precise identification of these two states requires
further analysis of more channels, such as $J/\psi\to K_SK^\pm\pi^\mp$
and $J/\psi\to K^+K^-\eta$.
Our PWA solutions have notable differences from those
presented in Ref.~\cite{Ablikim:2006hp} and
more recently in Ref.~\cite{Lees:2017ndy}.
We also report the most precise measurement of the branching fraction $B(J/\psi \to K^+K^-\pi^0)$.

\section{Acknowledgements}
The BESIII collaboration thanks the staff of BEPCII and the IHEP computing center for their strong support. This work is supported in part by National Key Basic Research Program of China under Contract No. 2015CB856700; National Natural Science Foundation of China (NSFC) under Contracts Nos. 11625523, 11635010, 11735014; National Natural Science Foundation of China (NSFC) under Contract No. 11835012; the Chinese Academy of Sciences (CAS) Large-Scale Scientific Facility Program; Joint Large-Scale Scientific Facility Funds of the NSFC and CAS under Contracts Nos. U1532257, U1532258, U1732263, U1832207; CAS Key Research Program of Frontier Sciences under Contracts Nos. QYZDJ-SSW-SLH003, QYZDJ-SSW-SLH040; 100 Talents Program of CAS; INPAC and Shanghai Key Laboratory for Particle Physics and Cosmology; German Research Foundation DFG under Contract No. Collaborative Research Center CRC 1044; DFG and NSFC (CRC 110); Istituto Nazionale di Fisica Nucleare, Italy; Koninklijke Nederlandse Akademie van Wetenschappen (KNAW) under Contract No. 530-4CDP03; Ministry of Development of Turkey under Contract No. DPT2006K-120470; National Science and Technology fund; The Knut and Alice Wallenberg Foundation (Sweden) under Contract No. 2016.0157; The Royal Society, UK under Contract No. DH160214; The Swedish Research Council; U. S. Department of Energy under Contracts Nos. DE-FG02-05ER41374, DE-SC-0010118, DE-SC-0012069; University of Groningen (RuG) and the Helmholtzzentrum fuer Schwerionenforschung GmbH (GSI), Darmstadt. This paper is also supported by the NSFC under Contract Nos. 10805053.

\bibliography{base}

\end{document}